\documentclass[12pt,preprint]{aastex}

\usepackage{epsfig}
\usepackage{graphicx}
\usepackage{epstopdf}
\usepackage{amsmath}

\usepackage{rotating}

\usepackage{natbib}
\begin{document}

\title{Discovery and Precise Characterization by the MEarth Project of LP 661-13, an Eclipsing Binary Consisting of Two Fully Convective Low-mass Stars}
\author{Jason A. Dittmann$^{1}$, Jonathan M. Irwin$^{1}$, David Charbonneau$^{1}$, Zachory K. Berta-Thompson$^{2,3}$, Elisabeth R. Newton$^{1,4}$, David W. Latham$^{1}$, Christian A. Latham$^{1}$, Gilbert Esquerdo$^{1}$, Perry Berlind$^{1}$, Michael L Calkins$^{1}$}
\affil{[1] Harvard-Smithsonian Center for Astrophysics, 60 Garden St., Cambridge, MA, 02138} 
\affil{[2] Torres Fellow, Massachusetts Institute of Technology, Cambridge, MA, 02138}
\affil{[3] University of Colorado, 389 UCB, Boulder, Colorado, 80309}
\affil{[4] NSF Astronomy and Astrophysics Postdoctoral Fellow, Massachusetts Institute of Technology, Cambridge, MA, 02138}

\begin{abstract}
We report the detection of stellar eclipses in the LP 661-13 system. We present the discovery and characterization of this system, including high resolution spectroscopic radial velocities and a photometric solution spanning two observing seasons. LP 661-13 is a low mass binary system with an orbital period of $4.7043512^{+0.0000013}_{-0.0000010}$ days at a distance of $24.9 \pm 1.3$ parsecs. LP 661-13A is a $0.30795 \pm 0.00084$ $M_\odot$ star while LP 661-13B is a $0.19400 \pm 0.00034$ $M_\odot$ star. The radius of each component is $0.3226 \pm 0.0033$ $R_\odot$ and  $0.2174 \pm 0.0023$ $R_\odot$, respectively. We detect out of eclipse modulations at a period slightly shorter than the orbital period, implying that at least one of the components is not rotating synchronously. We find that each component is slightly inflated compared to stellar models, and that this cannot be reconciled through age or metallicity effects. As a nearby eclipsing binary system where both components are near or below the full-convection limit, LP 661-13 will be a valuable test of models for the structure of cool dwarf stars.
\end{abstract}
\keywords{stars: fundamental parameters, stars: low-mass, stars: individual (LP 661-13), binaries: eclipsing, solar neighborhood}

\section{Introduction}
The M dwarf spectral sequence spans a large range of mass, from 0.6 M$_{\odot}$ at the earliest spectral types down to the main sequence turn off at approximately 0.08 M$_{\odot}$. This mass range spans important transitions in the physical structure of the interior of these stars. Notably, these stars transition to the fully convective regime midway through the spectral sequence, at $0.35 M_{\odot}$ \citep{structure_convection}. These transitions must be accurately captured in stellar models and their effect on the equations of stellar structure must ultimately be reflected in the temperatures and radii of these stars. However, the fundamental properties of low-mass stars remain a significant challenge for stellar structure models, particularly below 0.35 solar masses \citep{Torres_Structure,New_Baraffe_Models_2015}. 

Testing the mass-radius relation for low-mass stars is traditionally done through the study of eclipsing binary (EB) systems. Precise radial velocity measurements taken throughout the orbit are sensitive to the component masses of the system, while measurements of the eclipse depths and shapes are sensitive to the radii of the eclipsing stars. The best-observed, detached, double-lined EBs can provide measurements accurate at the $1\%$ level, allowing these systems to become strong tests of current stellar models \citep{Torres_2010_Review}. 

One of the closest and most well studied eclipsing binary systems is CM Draconis (CM Dra). CM Dra is an eclipsing M dwarf binary at a distance of 14.5 parsecs from the Sun \citep{CM_Dra_discovery,Lacy77}. CM Dra also contains a white dwarf at a wide separation from the M dwarf pair. As instrumentation and modeling have improved, the masses and radii of the two M dwarfs are now measured at the $0.5\%$ level \citep{Metcalfe96,Morales09}. Both stars in the CM Dra eclipsing binary are spectral type dM4.5 with masses of 0.23 and 0.21 solar masses and radii of 0.25 and 0.24 solar radii and orbit in a 1.7 day orbit \citep{Morales09}. The radii of these stars are inflated at the $5\%-7\%$ level \citep{Morales09}, and this remains a problem even with the latest stellar models \citep{New_Baraffe_Models_2015}. 

This problem is not restricted to a handful of systems. Taking only the most well-measured eclipsing binary systems in aggregate, low-mass stars tend to be \emph{inflated} in radius and \emph{cooler} in temperature than stellar models predict \citep{Torres_Structure}. The number of low mass stars with stellar radii measured through interferometry is low \citep{interferometry}, and often do not have any direct means to measure a precise mass. High precision measurements of stellar masses and radii for individual stars are obtained through measurements of eclipsing binary light curves and radial velocities (RVs). Photodynamical analyses of recently discovered triple systems (such as KOI-126, \citealt{carter2011} and Kepler-16 \citealt{Kepler16}) allow even more accurate physical parameters measurements than classical eclipsing binaries due to the presence of eclipses between all three members of the system. However, since the eclipse probability and the probability of detection are strong functions of orbital separation, these systems tend to be dominated by close-in binaries. This makes them more susceptible to the effects of tidal forces between the stars and makes them likely to be tidally locked, preventing the stars from spinning down over their lifetime. This effect makes it more likely for these systems to be magnetically active and to remain significantly magnetically active over their main-sequence lifetimes. If magnetic activity can significantly affect the interior structure of low-mass stars, then this can create an observational bias in the observed radii of these stars, especially if these effects are not accounted for in models.

One way around this problem is to search for low-mass eclipsing binary systems that have sufficiently long periods that both stars essentially evolve as ``single" stars, and collect enough high-quality data to constrain their physical parameters to sufficient accuracy to test existing stellar models. While difficult, several such systems have recently been discovered and characterized. The Kepler-16 system consists of a 41 day period eclipsing binary system orbited by a planet in a 229 day orbit, all exhibiting mutual occultations of each other \citep{Kepler16}. The presence of both stellar occultations and planetary transits in this unique system allowed for extremely precise physical parameters of this system to be measured. The secondary star in the system is a 0.202 M$_\odot$, 0.226 R$_\odot$ star with a mass and radius measured with sub-1\% precision. Our group has discovered a 41-day eclipsing binary, LSPM J1112+7626, consisting of two M-dwarfs. These stars have masses of $M_1 = 0.395$ M$_\odot$ and $M_2 =  0.275$ M$_\odot$ and radii of $R_1 = 0.382$ R$_\odot$ and $R_2 = 0.300$ R$_\odot$ \citep{Irwin_41day}. The masses of each component are measured with 0.5\% precision and the radii with 1.5\% precision. In addition to these long period systems, several other eclipsing binary systems with low-mass stellar components have recently been found with periods in the 5-20 day range \citep{2013ApJ...768..127S,2014A&A...572A..50G,2015MNRAS.451.2263Z}. Some of the components in these systems show radii that are consistent with stellar models, while some show significant radius inflation. Assessing the cause of the radius inflation phenomenon in low mass stars requires the discovery of sufficient numbers of these systems with sufficiently different physical characteristics (orbital separation, metallicity, etc) as well as accurate determinations of the mass and radius of each component. 

MEarth is an ongoing photometric survey of mid-to-late M dwarfs in the solar neighborhood (Distance, $D \lesssim 33$ pc), looking for low mass rocky planets whose periods may extend into the habitable zone of their star \citep{Nutzman,Berta_2013,jonathan_cool_stars}. The MEarth-North array in Arizona has been in operation since 2008, and a copy located in Cerro Tololo, Chile has been in operation since early 2014. By virtue of being designed to be sensitive to small planets transiting these stars, MEarth is also highly sensitive to eclipsing binary systems.

Here we present the discovery of an eclipsing binary system revealed during the commissioning phase of the MEarth South array. This system shows out of eclipse modulations due to star spots which change between observing seasons. Through long-term out of eclipse monitoring, we are able to assess the impact that transient starspots have on our ability to measure the radii of each component, which in turn allows us to more reliably probe the physical parameters of this system and assess our errors. We utilize multiple eclipse measurements with the MEarth telescopes as well as radial velocity (RV) measurements in order to constrain the masses and radii of both components to high accuracy and test existing stellar models. In section 2, we detail the MEarth-South array, the discovery, and the follow-up observations of this system. In section 3, we present a joint analysis of the RV and photometric data and constrain the physical parameters of the system. In section 4, we discuss the implications of these measurements in regards to existing theoretical stellar models. 

\section{Observations}

\subsection{The MEarth-South Observatory}
MEarth-South, like its Northern counterpart, consists of eight f/9 40 cm Ritchey-Chr\'etien telescopes on German equatorial mounts. The telescopes are located at Cerro-Tololo International Observatory (CTIO) in Chile \citep{jonathan_cool_stars}. The telescopes are automated and take data on every clear night. Each telescope is equipped with a photometer that utilizes a 2048 $\times$ 2048 pixel CCD with a pixel scale of approximately 0.84\arcsec / pixel and a Schott RG715 glass filter with an anti-reflection coating. For information about our filter curve and photometric system, see \citet{mearth_photometry}. We expect that the MEarth-South photometric system is slightly different from the North, and that the results in \citet{mearth_photometry} might show minor differences with the Southern array.  The MEarth-South CCDs are e2v CCD230-42 devices with a NIR optimized coating with fringe suppression. We operate the MEarth cameras at $-30^{\circ}$ C, and before each exposure we pre-flash the detector. This increases the dark current (which we subtract off), but it also eliminates persistence from the previous exposure.

We gather sky flat field frames at dawn and at dusk. Since MEarth uses German equatorial mounts, we must rotate the telescopes by 180 degrees relative to the sky when crossing the meridian. Therefore, we take two sets of flat fields, taking adjacent pairs of flat fields on opposite sides of the meridian. This procedure also allows us to average out large-scale illumination gradients from the Sun and the Moon. This scattered light also concentrates in our detectors in the center of the field of view. The amplitude of this scattered light effect is approximately 5\% of the average value of the sky across the CCD. In order to correct for this, we filter out the large scale structure from our combined twilight flat field and use the residual flat field to track small scale features such as inter-pixel sensitivity and dust shadows on the detector. We derive the large scale flat field response from dithered photometry of dense star fields. 

We measure the nonlinearity of the MEarth detectors using a dedicated sequence of dome flats. At all count levels, the MEarth CCDs have a slightly nonlinear conversion of photoelectrons to data number. This nonlinearity increases from 1\%-2\% at half the detector full well to 3\%-4\% near saturation. We correct for this effect as part of the general MEarth data reduction pipeline. Our exposure times are set to avoid surpassing 50\% of the detector's full well in order to minimize this effect. 

The MEarth-South target list is designed to be similar to the northern array's target list \citep{Nutzman}. MEarth-South observes a selection of nearby M dwarfs believed to be within 33 parsecs (pc) of the Sun and with a stellar radius $R < 0.33 R_\odot$. The completeness of the target list and the number of targets with measured parallaxes is much lower in the South than in the North because the northern target list has been assembled from a much wider array of sources. No analogous compilation existed for the Southern hemisphere. For this reason, and also the availability of follow-up resources, the majority of our targets lie above $-30^{\circ}$ DEC, rather than being distributed uniformly across the Southern sky ($\delta < 0^{\circ}$). Typically there is only one target in each MEarth-South 26\arcmin $\times$ 26\arcmin$ $ field of view, with the exception of wide multiple systems and occasional unrelated asterisms. Each field is targeted individually and is selected to contain sufficient comparison stars to enable high precision relative photometry. 

Each clear night MEarth observes a set of visible target stars at a cadence of approximately 20-30 minutes. This ensures that we obtain at least two in-transit data points for a typical one hour duration transit. If the real-time reduction software detects a potential transit (or eclipse) in-progress, standard cadence observations of other targets on that telescope are interrupted in order to perform high cadence follow-up of the potential event-in-progress (see \citealt{Berta_2013} for a description of the MEarth trigger). Normal operations are resumed when the event is deemed to be over or spurious and the flux from the star has returned to its normal level.

\subsection{Initial Detection and Follow-up}
\begin{figure}
%\begin{singlespace}
\centering
\includegraphics[width=0.75\linewidth]{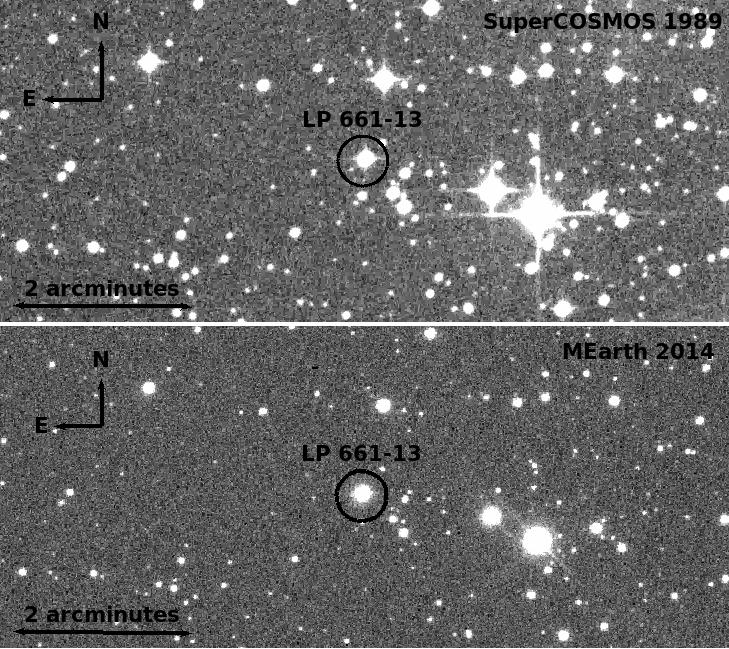}
\caption{Finding chart for LP 661-13. The top image is archival data taken by the SuperCOSMOS survey from 1989. The bottom image is a MEarth image taken in 2014. The circle does not represent the MEarth aperture; its size is selected for clarity. The SuperCOSMOS image indicates that there is a faint background star that currently sits in the MEarth aperture for LP 661-13. We address this third light contamination later in the text.}
\label{finding_chart}
%\end{singlespace}
\end{figure}

A primary eclipse of LP 661-13 (06$^h$56$^m$18$^s$.95, -08$^{\circ}$35\arcmin46.1\arcsec; alternate names NLTT 17194, 2MASS J06561894-0835461) was first discovered on 28 January 2014, on the second night of commissioning observations of the MEarth-South array (see Figure \ref{finding_chart} for a finding chart for this system). Due to being in commissioning phase, the MEarth trigger  was not yet being used. Therefore, only 6 in-eclipse data points were taken during the event. Subsequent eclipses were observed on the nights of February 16, 2014 (primary eclipse) and February 23, 2014 (secondary eclipse), and utilized the MEarth trigger. These events, combined with out-of-eclipse monitoring in the intervening nights, allowed us to unambiguously identify the orbital period as approximately 4.704 days. Armed with orbital period of the system, we began strategic targeting of radial velocity observations and subsequent eclipse measurements.

We obtained radial velocity measurements with the Tillinghast Reflector Echelle Spectrograph (TRES) on the 1.5 m Tillinghast reflector located at the Fred Lawrence Whipple Observatory (FLWO) on Mt. Hopkins, Arizona. We used the medium (2.3\arcsec) fiber, which yields a resolving power of $R = \frac{\lambda}{\Delta \lambda} = 44,000$. We obtained 14 observations of LP 661-13 with TRES, with 8 exposures taken in the 2014 observing season (18 February 2014 \-- 28 March 2014) and 6 taken in the 2015 observing season (27 December 2014 \-- 06 January 215). Total exposure times were 1 hour per observation, except for one exposure which was was 40 minutes. We obtained Thorium-Argon (ThAr) wavelength calibration exposures both before and after each science exposure. These wavelength calibration exposures were obtained with the same fiber as the one used to take the data on the target itself. 

Follow-up photometric observations of both primary and secondary eclipses were performed with the MEarth-South telescopes. We observed 11 unique eclipse events with 6 different telescopes. In addition to the observations of individual eclipses, we monitored the brightness of LP 661-13 out of eclipse on each clear night from CTIO in order to assess the variability of the system due to the stellar rotation of each component bringing star spots into and out of view. We have obtained 9173 individual observations of LP 661-13, with 5098 observations being in-eclipse measurements and 4075 out-of-eclipse measurements.

\section{Analysis}
\subsection{Spectroscopy}
We reduced our TRES spectroscopic data using the procedure of \citet{tres_reduction}, the standard pipeline for radial velocity analysis with TRES. Once these data are reduced, we use a two-dimensional cross correlation algorithm \texttt{TODCOR} \citep{todcor94}, which uses user-given templates to match each component in the system, derive their radial velocities, and measure their light ratio in the observed spectral bandpass. For both components we used a single epoch observation of Barnard's star (GL 699) taken on 18 April 2011 as our template. Barnard's star has a spectral type of M4 \citep{1991ApJS...77..417K}, and is suitable for both stars in this analysis. We also performed a reduction using the later type Wolf 359 as a template but found Barnard's star to yield more precise results. We correlated our spectra using the wavelength ranges of 8270-8430\AA  $ $  (aperture number 41). This region contains molecular features, which produce strong correlations and are able to robustly identify both components in the spectrum. 

First, we fit for the radial velocities for both components while letting the light ratio of the two components be a free parameter in the \texttt{TODCOR} model. We let the light ratio vary between each observation. The light ratio is determined by the ratio of the strengths of the features in spectrum for each component, after matching to the stellar template (in this case, Barnard's star). Once we obtained an initial solution for each epoch, we averaged the fitted light ratio for each epoch. We then fixed the light ratio to this value and re-derived the best fit radial velocities for each component. We find a best fitting light ratio of $L_2 / L_1 = 0.434 +/- 0.025$. Using the derived physical parameters for our system, the temperature scale of \citet{mann15}, and the models of \citet{Stellar_Models_flux_ratio}, we predict a light ratio of $L_2 / L_1 = 0.406$ for this system at these wavelengths, consistent with what we measure through our high resolution spectra. We do not use this light ratio when fitting our photometry.

\begin{figure}
%\begin{singlespace}
\centering
\includegraphics[width=0.75\linewidth]{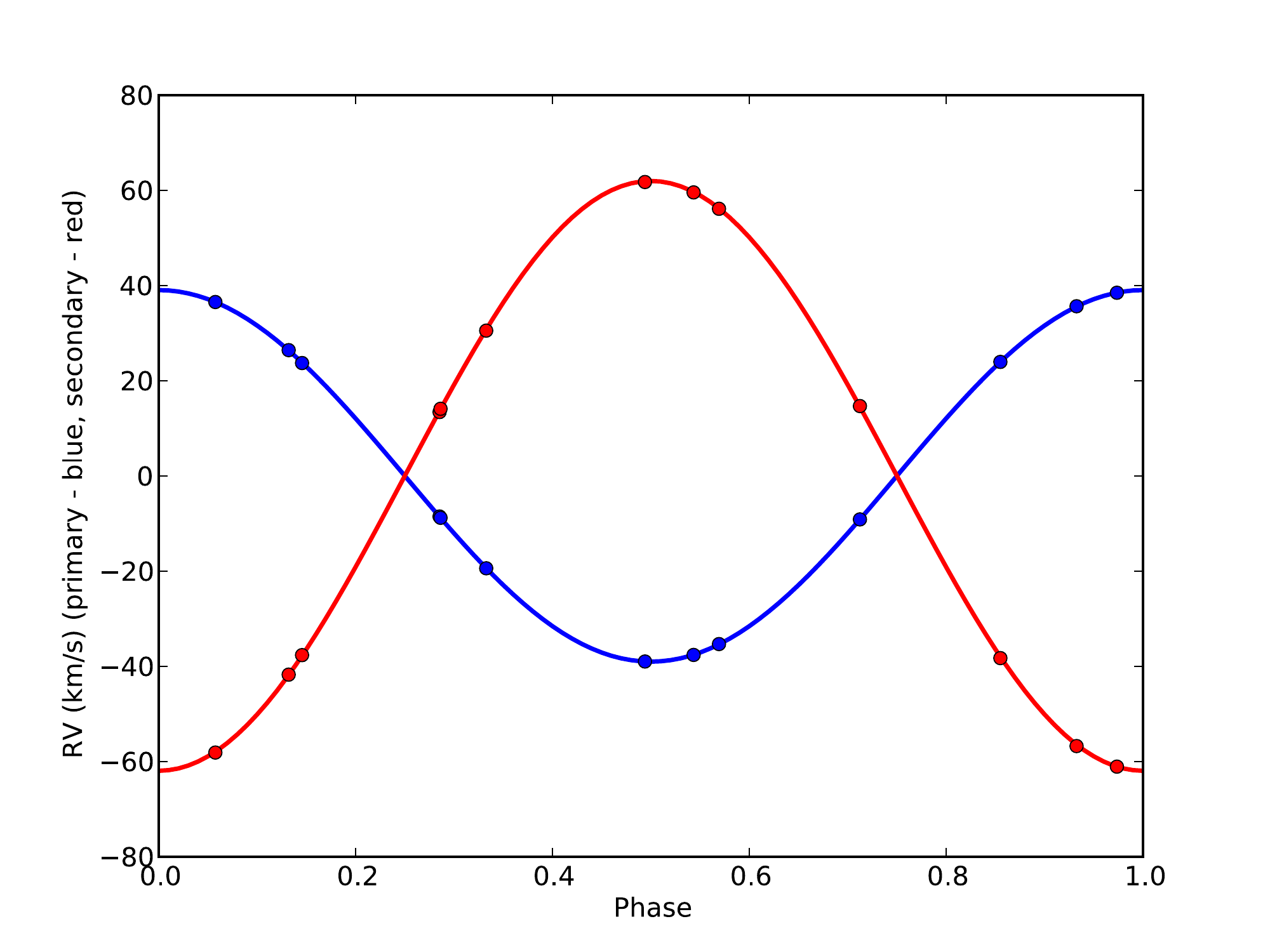}
\includegraphics[width=0.75\linewidth]{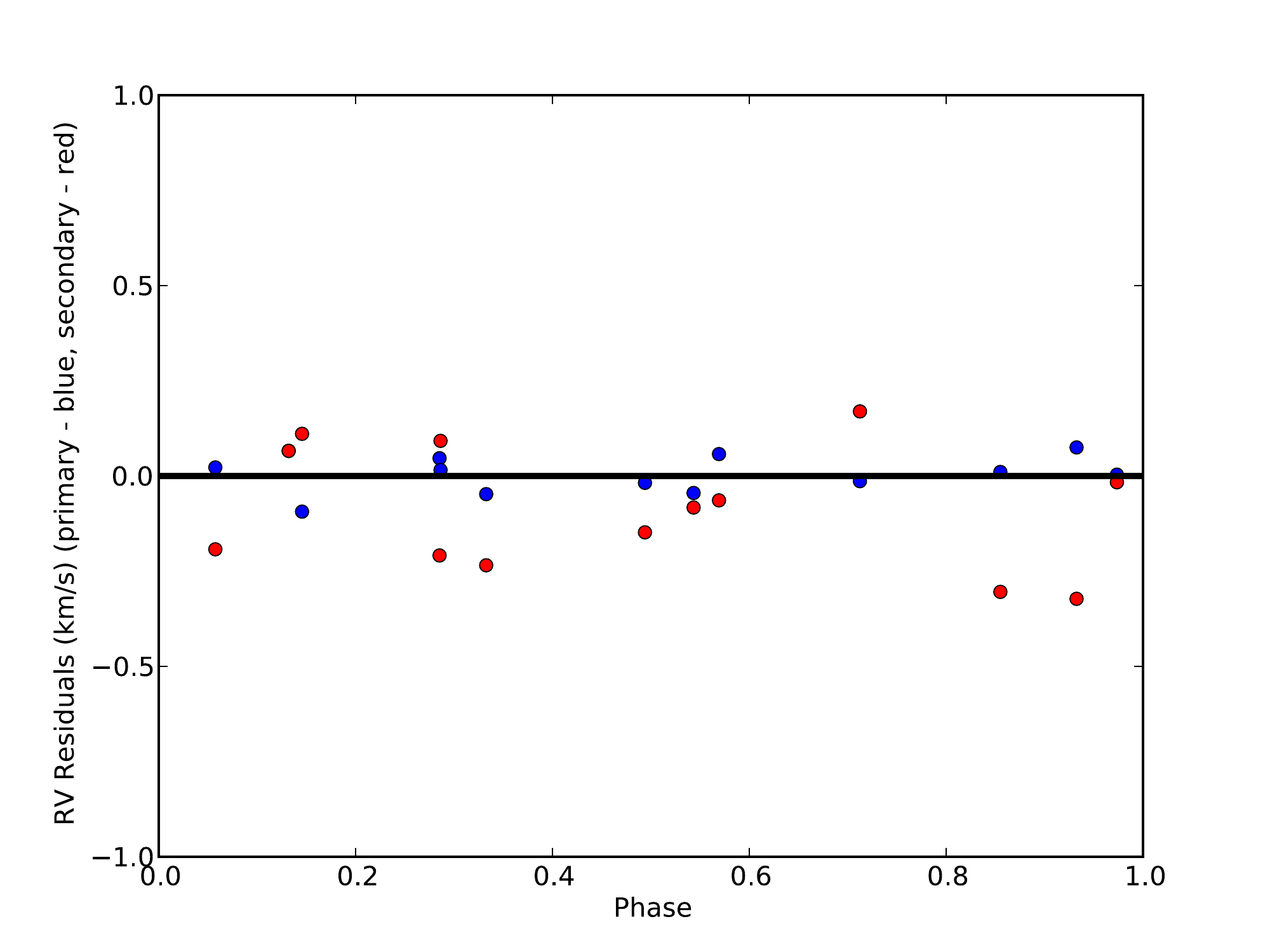}
\caption{Top: Radial velocity signal from each component of LP 661-13. The primary component is in blue and the secondary component is in red. We find a primary mass of $M_1$ = $0.3057 \pm 0.0032$ $M_\odot$ and $M_2$ = $0.1930 \pm 0.0014$ $M_\odot$) orbiting with a period of P = $4.7043518^{+0.0000017}_{-0.0000014}$ days. Bottom: Residuals from the fit. We find an RMS precision of 0.05 km s$^{-1}$ for the primary component and 0.20 km s$^{-1}$ for the secondary component.}
\label{RV_plot}
%\end{singlespace}
\end{figure}

We report the radial velocities from this analysis in Table \ref{RV_table} and show a plot of the heliocentric radial velocities in Figure \ref{RV_plot}. We omit one observation taken on BJD 2456743 as we could not resolve both components in the spectrum due to their proximity in velocity space for this epoch and the likelihood that the primary peak would be systematically skewed in velocity due to the unresolved secondary peak in the cross correlation. We estimate the uncertainty of our velocities (based on the standard deviation of our residuals) as 0.05 km s$^{-1}$ for the primary and 0.20 km s$^{-1}$ for the secondary. We assume a barycentric velocity for Barnard's Star of $-110.506 \pm 0.035$ km s$^{-1}$ \citep{barnards_star_RV}, and derive a barycentric velocity for this system of $-0.009 \pm 0.037$ km s$^{-1}$, including both our internal uncertainty in the barycentric velocity of the system and the error in the determination for Barnard's star (see modeling section \ref{model} for a description of how we obtained this quantity.).

\subsection{Photometry}
We measure stellar positions using a method similar to \citet{I85}. We estimate the local sky background by binning each image in 64 $\times$ 64 pixel blocks and measure the peak of the histogram of the intensity of these pixels. This lower-resolution map is then interpolated to measure the background level anywhere in the image. Sky background is estimated with a sky annulus between 18 and 24 pixels away from the stellar photo center. We measure stellar locations from the intensity weighted first moment (i.e. the centroid) of the star. 

We measure the total flux using a 6 pixel ($\approx 5.04\arcsec$) aperture radius. We weight pixels that lie partially within this circular aperture by the fraction of the pixel that lies within the ideal circular aperture. We also adopt an aperture correction to correct for the stellar flux that falls outside of our aperture. The typical size of an aperture correction is 0.04 magnitudes, but can vary from night to night depending on atmospheric conditions (predominantly seeing). In Table \ref{Photometry_table}, we provide the corrected photometry for LP 661-13 across all telescopes and both seasons of data.

There is a marginal Roentgan Satellite (ROSAT) detection in the ROSAT faint source catalogue \citep{rosat_faint}. The potential source has a count rate of $3 \pm 1 \times 10^{-2}$ counts per second. This corresponds to a flux density of approximately $1.70 \pm 0.56 \times 10^{-13}$ ergs cm$^{-2}$ sec$^{-1}$. While this may be interesting to assess the significance of the X-ray emission with stellar activity and radius inflation, further X-ray data to confirm this potential detection are needed.

\subsection{Astrometry}
With two seasons of photometric data from the MEarth observatory, we are able to utilize the MEarth astrometric pipeline to measure the trigonometric distance to LP 661-13. The astrometric pipeline is described in detail in \citet{Dittmann_parallax}, but we summarize it here. We eliminate nights for which the full-width half-max for the image is greater than 5 pixels (approximately 3.5 arcseconds), or the average ellipticity of the target stars is greater than 0.5 (typically due to wind shake). We also eliminate images for which the pointing on the image is more than 15 pixels discrepant from the master MEarth image. We fit each image with a linear model in both the x and y coordinates, which allows for translation, rotation, shearing and pixel scale variations. We also include a separate constant offset between data taken on opposite sides of the meridian, to accommodate the image flip that results from the telescope's german equatorial mount.

We choose reference stars that lie within 600 pixels of the target star to avoid higher order effects at the edges of the CCD. We fit our astrometry iteratively, first fitting the linear plate model to each frame and then fitting for the proper motion and parallax of each star until convergence. Internal errors are estimated through a residual permutation. This provides the advantage that it preserves time-correlated noise in our error assessment but tends to be a weaker way to estimate errors for series with few data points. However, we have over 9000 data points and do not believe our astrometric error bars are underestimated.

\begin{figure}
%\begin{singlespace}
\centering
\includegraphics[width=0.75\linewidth]{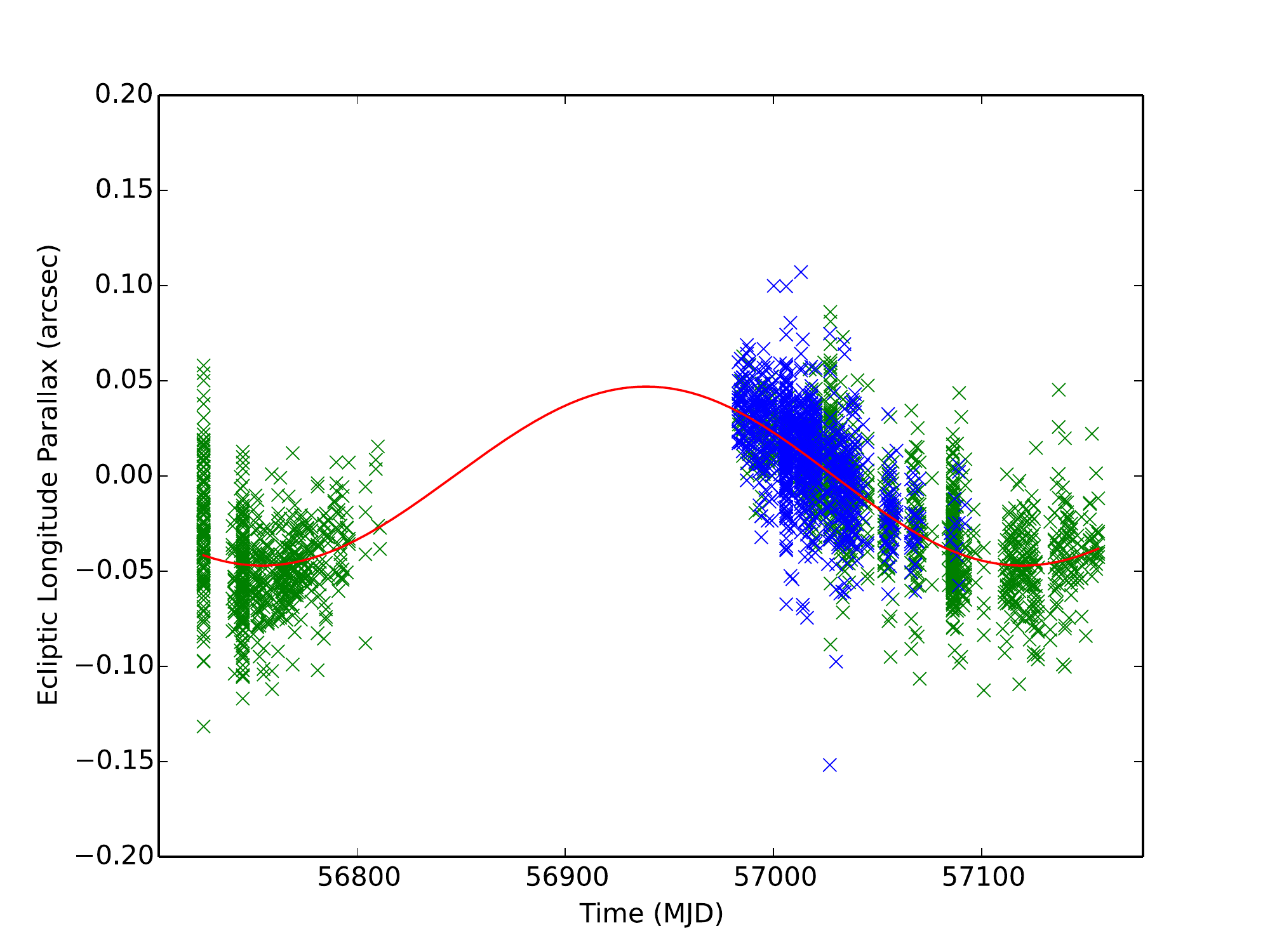}
\includegraphics[width=0.75\linewidth]{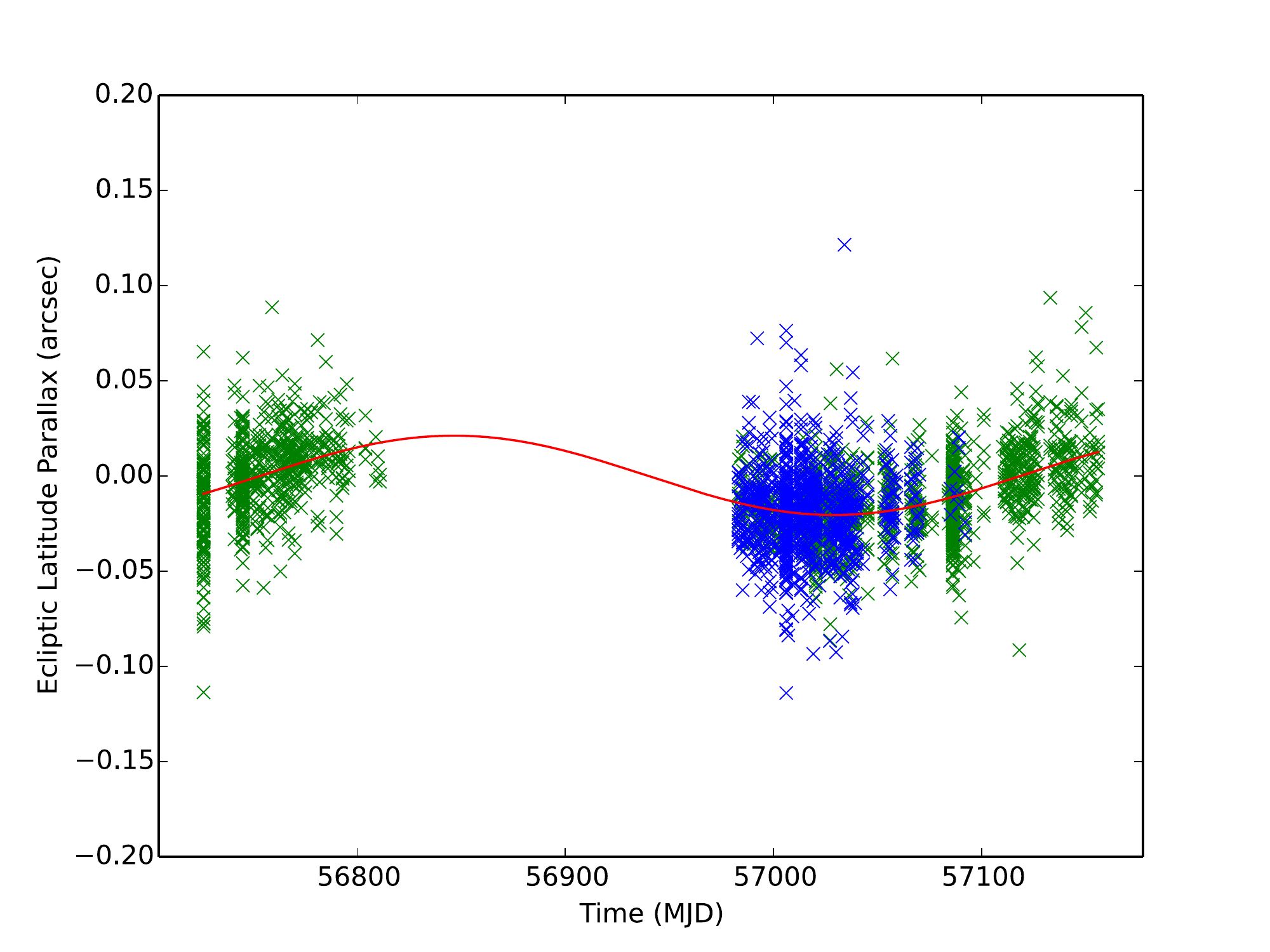}
\caption{Trigonometric parallax signal from MEarth-South images in ecliptic longitude (left) and ecliptic latitude (right) for LP 661-13. We follow the procedure of \citet{Dittmann_parallax}, as MEarth-South and MEarth-North were built to be nearly-identical arrays. We find a parallax of $\pi = 40.2 \pm 2.0$ mas, which corresponds to a distance of $24.87 \pm 1.3$ parsecs.}
\label{parallax_figure}
%\end{singlespace}
\end{figure}

We perform our astrometric analysis for only the telescope which has the largest number of data points over the largest time baseline. This time baseline ensures we can resolve any degeneracies between the proper motion and the parallax of the star. Using only one telescope eliminates possible systematics with combining data from multiple telescopes. Additionally, observations from other telescopes were targeted observations of eclipses and are of little use for astrometric analysis, which require data to be taken over the Earth's entire orbit. We show our astrometric time series in Figure \ref{parallax_figure}, and measure a trigonometric parallax of $\pi = 40.2$ milliarcseconds (mas) with an internal error of 1.0 mas. However, since MEarth-South astrometry has not been benchmarked to a sample of stars with previously measured parallaxes (unlike MEarth-North), we inflate this error bar by a factor of 2 to be conservative. This corresponds to a distance of $D = 24.87 \pm 1.3$ parsecs (pc). We do not calculate a correction from relative parallax to absolute parallax, as this effect is smaller than our errors. We find proper motions of $75.1 \pm 2.0$ mas yr$^{-1}$ in the RA direction and $-227.3 \pm 2.0$ mas yr$^{-1}$ in the DEC direction, where we have also inflated our internal error bars in the same manner as for the parallax amplitude. These values are consistent with the proper motion of $76.9 \pm 5.5$ mas yr$^{-1}$ in the RA direction and $-219.6 \pm 5.5$ mas yr$^{-1}$ previously reported by \citet{proper_motion}

\section{System Modeling}
\label{model}
We model this system following a similar procedure as \citet{Irwin_41day}. While for LSPM J1112+7626 \citet{Irwin_41day} notes that the system is eccentric and therefore the photometric and spectroscopic solutions are highly interrelated, for LP 661-13 we find no detectable eccentricity, and so the analysis is made simpler. We model the system using the light curve generator from \texttt{JKTEBOP}\footnote{See http://www.astro.keele.ac.uk/jkt/codes/jktebop.html for the original \texttt{JKTEBOP} code} (\citealt{jktebop1}; see \citealt{jktebop2} for the most recent version). \texttt{JKTEBOP} is based on the eclipsing binary program \texttt{EBOP} \citep{ebop}. We use a modified version derived from \citet{Irwin_41day} which computes the integrals analytically using the method of \citet{mandel_and_agol} and its erratum. 

The out of eclipse light curve modulation indicates that there is significant spot activity on the surface of at least one component of the system, possibly both. This complicates the analysis as starspots have the capability of causing systematic errors in the measurement of the stellar radii as well as the surface brightness ratio of the two components, which is derived from the eclipse depths. Spots occulted during the eclipse reduce the observed depth of the eclipse relative to a non-spotted star, since a relatively dimmer portion of the star is being occulted. Conversely, the presence of non-occulted spots systematically increases the eclipse depth as the eclipse is preferentially obscuring brighter portions of the star. In reality, it is likely that both occulted and unocculted spots exist in any given eclipse. It is extremely difficult to infer the true spot distribution of a system, except in cases where eclipse deviations (i.e. spot crossing events) are readily distinguished in the light curve.  Out of eclipse modulations can help assess spot coverage, but are sensitive to only asymmetric distribution of star spots. A uniformly spotted star would show no variation at all because a starspot rotating out of view would be replaced by an identical spot rotating into view. Identifying, or failing to identify, multiple frequencies in the out of eclipse modulations can also help determine whether the starspots are concentrated on one star or if both components display significant asymmetric spot coverage, although if only one period in the out-of-eclipse modulation is detected it is impossible to determine on which component the signal originates.

We adopt the spot model from \citet{Irwin_41day}, which assumes that the out of eclipse modulations are solely due to starspots, and assumes a simple sinusoidal form for the modulations. The functional form for these modulations is:

\begin{equation}
\frac{\Delta L_i}{L_i} = a_i \ \mathrm{sin} \left( \frac{2\pi F_i t}{P} \right) + b_i \ \mathrm{cos} \left( \frac{2\pi F_i t}{P} \right) - \sqrt{a_i^2+b_i^2}
\end{equation}

where $t$ is the time, $L$ is the light from the star, $i$ denotes each component of the system, $a_i$ and $b_i$ are the constants expressing the amplitude and the phase of the out-of-eclipse modulation, and $F_i$ is the ratio of the rotational frequency of the star to the orbital frequency.  

In the present case, we find only one significant period for the out of eclipse modulations. However, since the observed modulation has a period close to the orbital period of the system, it is possible that both stars have similar periods and that they are close to a tidal locking scenario. This means that we cannot be sure on which component the starspots originate, and we must account for this systematic uncertainty. Unfortunately, fitting a spot model on both components is degenerate, given the limited information available from a single modulation derived from total integrated light measurements. In order to assess the full range of possible physical parameters that are consistent with the data, we fit a series of models with different spot parameters, which we describe below.

We fit both the spectroscopic RV data and the photometric light curve (both in and out of eclipse) simultaneously. We give the aggregate RV data set the same statistical weight as the entire photometric data set, despite the latter having thousands more individual data points. This is largely irrelevant, as the system is not eccentric and therefore the RV data and the photometric data are sensitive to independent system parameters. We assume here that there is no third light in the system; in Section 4.1 below, we show that third light would not significantly corrupt our estimates of the system parameters.

We fit two different models, one where there are starspots on only the primary component of the system, and one where the spots are concentrated on solely the secondary component of the system. We also fit the data from the 2014 season and the 2015 season independently, and allow the starspot modulation to change between seasons. Through this method, we aim to explore the possible effect that starspot contamination can have on our inference of the stellar radii. We explore this parameter space using the \texttt{emcee} code \citep{emcee}, which is a python implementation of the Affine Invariant Markov Chain Monte Carlo sampler. Each model is initiated with 100 walkers in a Gaussian ball located at an approximate solution. We run the chain for 50000 steps, and discard the first 5,000 steps to allow the solution to ``burn-in".  We use uninformative (unbound, uniform) priors for all of our model parameters, except for the limb darkening law. Quadratic limb darkening parameters are initiated with the model of \citet{claret2000} for each component utilizing an approximate temperature  (T$_{\textrm{eff}}$ = 3700K, log($g$) = 5.0) and the Cousins $I$ band filter. Each coefficient is allowed to vary freely, but not exceed a 10\% deviation from this theoretical value. This allows the model to adjust for differences in the actual star compared to the theoretical parameters as well as for the slight differences between our bandpass and $Cousins$ $I$ without venturing into physically implausible parameter space. Letting the limb darkening parameters float in this way allows us to explore our prior, and we do not contend that our results are a physical measurement of the actual limb darkening parameters for these stars. 

In Table \ref{model_table} we list the physical parameters of our model (not including the normalizations for each telescope and each eclipse), and in Tables \ref{results_table_prim} and \ref{results_table_sec} we show the resulting best fit model and the 16th and 84th percentile (approximately $1\sigma$) for each parameter for each season. We are able to fit our measured light curves equally well regardless on which star we place starspots. However, if we place the starspot signal on the secondary component, the out of eclipse model parameters vary significantly between observing seasons, whereas if the starspot signal is originating on the primary component, the parameters are stable between seasons. Particularly, F, the ratio between the orbital and rotational frequency has different values for each season if the secondary component is responsible for the star spot modulation signal. Since the rotation period of the secondary star should not change (and the effect of differential rotation for star spots at different latitudes is small for M dwarfs \citealt{differential_rot_davenport}), we conclude that the starspot signal cannot be originating on the secondary component and we will utilize the model for which starspots are located on the primary component for the rest of this paper. We note, however, that this choice does not significantly affect the values of the masses and radii of the components but does change the uncertainty. While the light curves themselves are most directly sensitive to the sum of the component radii and their ratio, with our data we are able to separate these two variables and measure the individual component radii themselves.

The base spottedness and the fraction of eclipsed spots model parameters are unconstrained by the data, and allowed to float to allow us to explore parameter space and assess our total uncertainty in our derived physical parameters. In order to assess possible systematic uncertainties associated with these model parameters, we also run MCMC chains holding these parameters fixed at 0.0, 0.25, 0.50, 0.75, and 1.0. We find that these parameters typically induce a $1.5$\% variation in the uncertainty of the radii of the component stars, and a negligible effect on the mass of the stars (which is primarily constrained with the RV data). While this systematic error is partially explored by letting these model parameters float in the chain, there will likely be some residual systematic error not captured in our MCMC runs. 

\begin{figure}
%\begin{singlespace}
\centering
\includegraphics[width=0.65\linewidth]{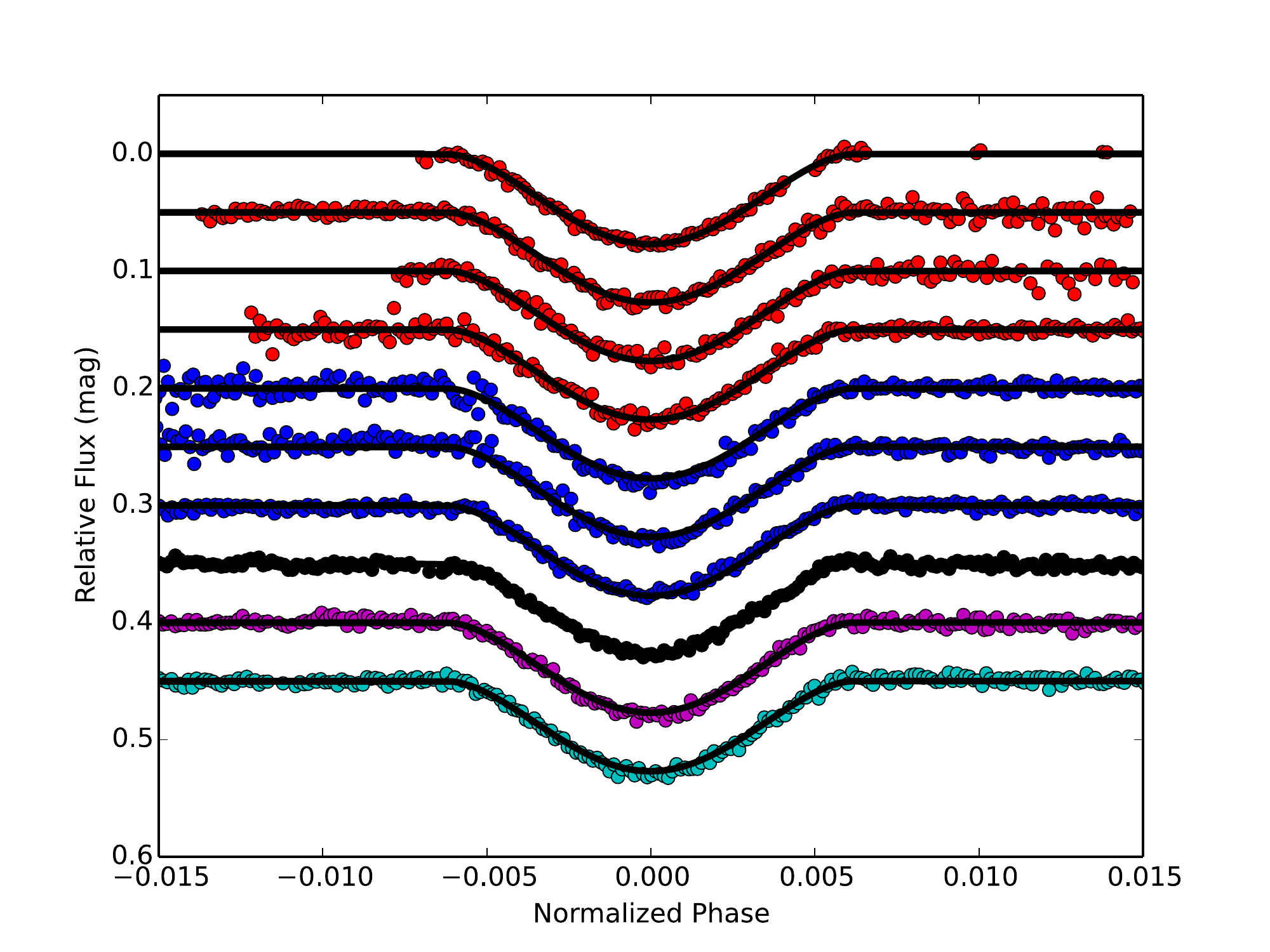} \\
\includegraphics[width=0.65\linewidth]{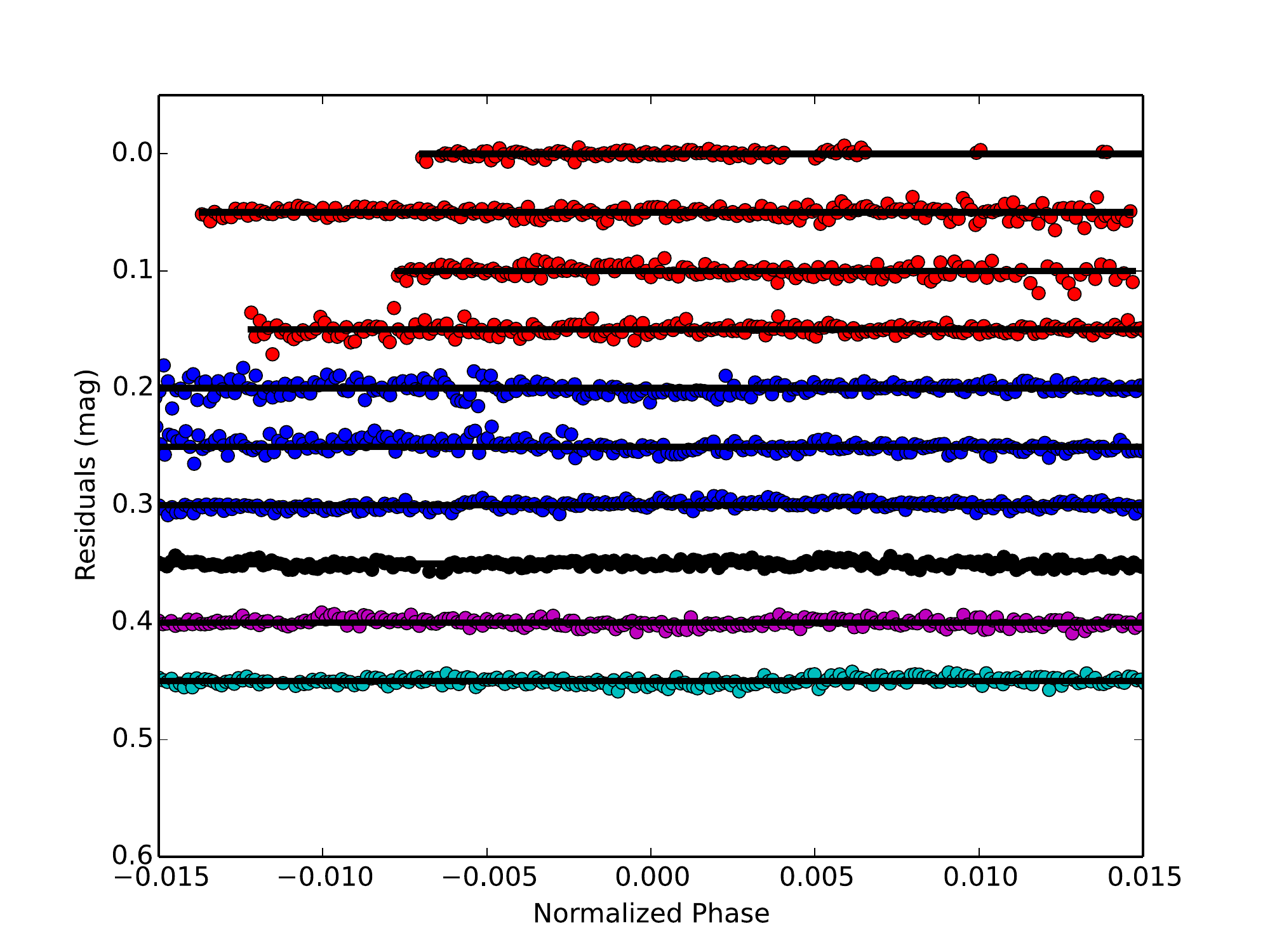}
\caption{Photometric data of all primary eclipses (offset for clarity) from the 2014 - 2015 observing seasons and our model with starspots located on the primary component. We have corrected the data for normalization, meridian offsets, and common mode, which are known systematics in the MEarth data. Residuals are located in the lower plot. Each color represents data taken from a different MEarth-South telescope.  }
\label{Primary_Eclipses}
%\end{singlespace}
\end{figure}

\begin{figure}
%\begin{singlespace}
\centering
\includegraphics[width=0.65\linewidth]{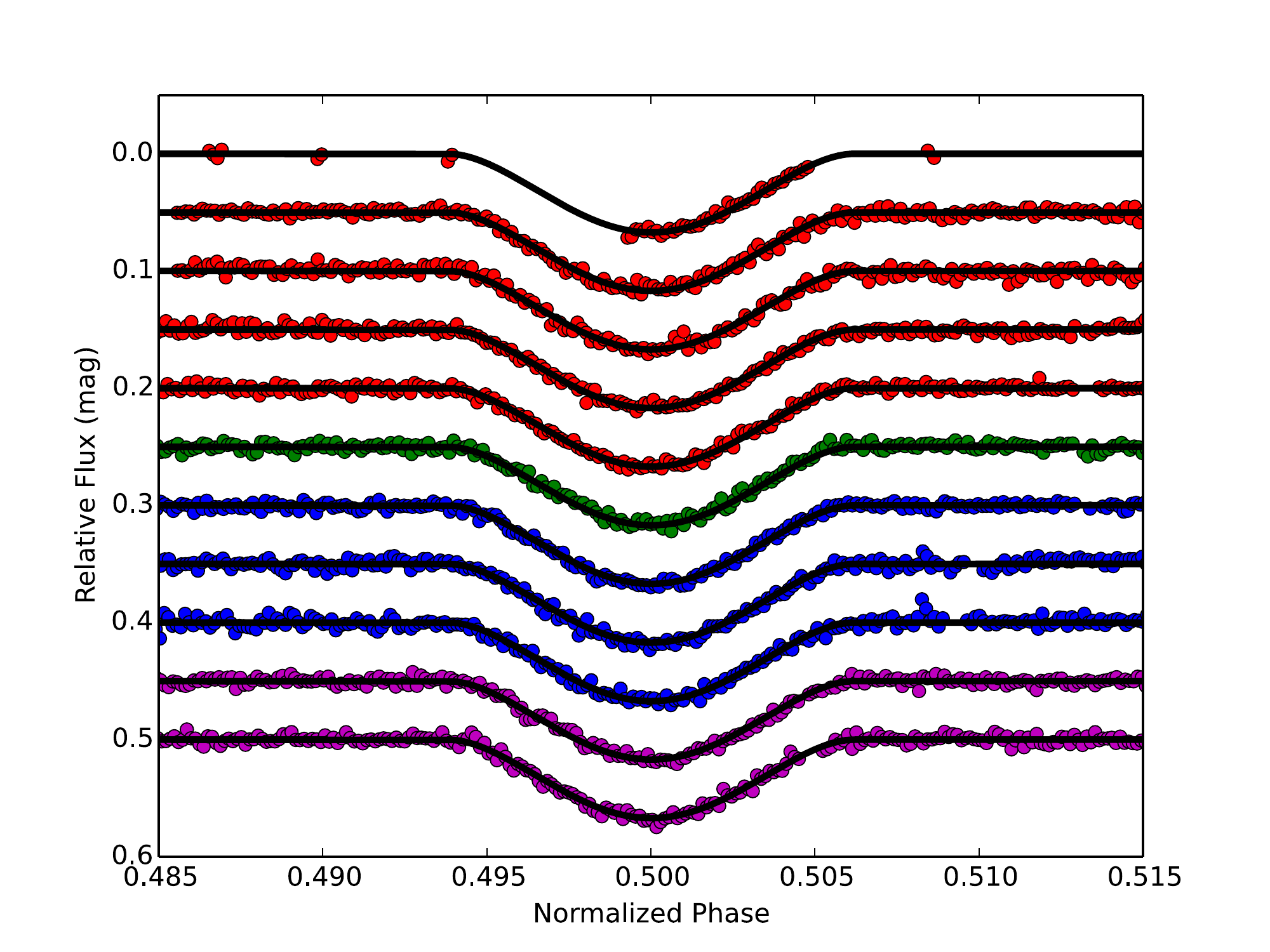} \\
\includegraphics[width=0.65\linewidth]{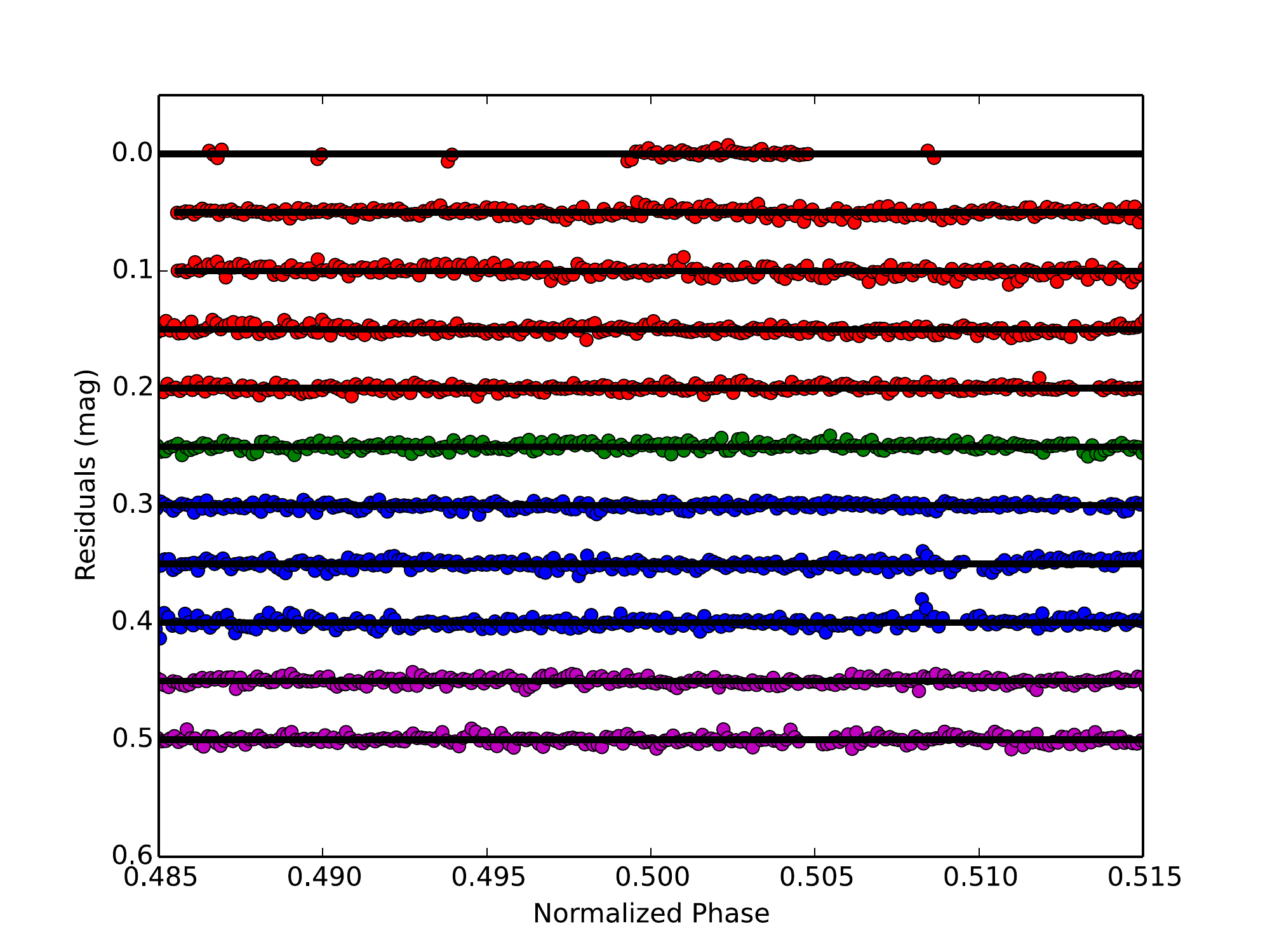}
\caption{Photometric data of all secondary eclipses (offset for clarity) from the 2014 - 2015 observing seasons. The partial secondary eclipse was data taken before we knew the period of the LP 661-13 system. MEarth-South automatically detected an event in-progress and began collecting high cadence follow-up observations until the event ended before resuming normal operations. The colors correspond to the same telescopes as in Figure \ref{Primary_Eclipses}. Residuals from the fit are shown in the lower panel.}
\label{Secondary_Eclipses}
%\end{singlespace}
\end{figure}

For our final solution, we fit both seasons jointly, requiring the same physical parameters across seasons but allowing the starspot model to change (but not change components). We have also fixed the fraction of eclipsed spots ($f_1$) to 1.0 and the base spottedness of the star ($f_2$) to 0.0, as it would take unlikely spot distribution on the surface of the star to significantly change these parameters. See \citet{Irwin_41day} for a detailed investigation into the role that these parameters play in affecting measured masses, radii, and brightness temperatures. We also fix the orbital eccentricity at 0, as we see no evidence for orbital eccentricity in this system from our eclipse timings (e $< 4\times10^{-4}$, 95\% confidence). In Table \ref{Final_Model_Params}, we list the final model parameters for our joint fit of both seasons of data assuming starspots distributed on the primary component and in Table \ref{Final_System_Parameters_table} we list the physical parameters, including component masses and radii, for the LP 661-13 system. In Figures \ref{Primary_Eclipses} and \ref{Secondary_Eclipses} we show each individual primary and secondary eclipse as well as their residuals with this model.

\subsection{Third Light}
Archival SuperCOSMOS images (see Figure \ref{finding_chart}) indicate that there is a faint background star currently in the MEarth aperture for LP 661-13. This third light can potentially limit our ability to measure the fundamental parameters of each component in this system. The SuperCOSMOS catalogues measures this background star to be 5.3 magnitudes fainter than LP 661-13 in the 103a-E red plate bandpass. Unfortunately, we do not know the color of the background star and therefore do not know the magnitude difference in the MEarth bandpass. While it is unlikely that the background star is as red as LP 661-13, we can measure the maximum possible effect that third light contamination has on determination of LP 661-13's physical parameters by repeating our analysis with third light fixed (arbitrarily) at 1\% of the total light of the system. In Tables \ref{Final_Model_Params} and \ref{Final_System_Parameters_table} we list the system parameters jointly fitting both seasons of data for third light, $L_3$ = 0 and 0.01 for models with starspots only on the primary stellar component, and for starspots only on the secondary stellar component. The most likely physical scenario is one where the starspots are concentrated on the more luminous primary and that third light is insignificant for this system. We find that with 1\% third light contamination, the radius of each component of LP 661-13 is affected at the $0.5$\% level, and therefore we do not believe third light to be a significant concern for this system.

\section{Discussion}
We have measured the masses of the primary and secondary components of LP 661-13 to a precision of 0.27\% and 0.17\% and the radii to 1.0\% and 1.4\%, respectively. At the photon noise limit, the ratio of the standard deviation of the residuals of our spectroscopic orbit scales inversely with the light ratio \citep{tres_photon_noise}. However, in our case, the ratio of our residuals is 0.25, significantly lower than our measured light ratio $L_2/L_1 = 0.434 \pm 0.025$, implying a significant red noise component in one or both component RV measurements. We further find that a systematic noise floor, present in both components and of the same magnitude, is unable to account for the internal error bars of our component masses. We do not believe this excess noise significantly affects our results, and we note that our determination of the stellar radii of both components is more uncertain than the masses.

The error in the radii is dominated by the systematics associated with the starspots, and further seasons of data will aid in reducing these errors. Particularly, if the starspot signal changes significantly between seasons this will aid our ability to break the degeneracies of this model. In future seasons, LP 661-13 will achieve wider separation with the background star, allowing us to measure its color and assess its third light contribution. 

In a previous work \citep{mearth_photometry}, we have shown that the absolute $K_s$ magnitude and the $MEarth - K_s$ color of a mid-to-late M dwarf can be an indicator of the star's metallicity. Here, we have measured the light ratio of the both component stars in the MEarth bandpasses, allowing us to separate the $MEarth$ magnitude into magnitudes for each star. While we don't have a similar measurement in the $K_s$ band, we can use our measured masses and the mass - luminosity relation from \citet{Delfosse} to separate the $K_s$ magnitude into the magnitudes of each component and our trigonometric parallax to convert to an absolute magnitude. We estimate the metallicity of the primary star is [Fe/H] = 0.0 and the metallicity of the secondary star is [Fe/H] = -0.13, with a precision of approximately $0.1$ dex, although this estimate inherits the precision and biases of the relation in \citet{Delfosse}. We do not believe that the stars in this system formed from different metallicity bulk material, and that the difference in our two estimates are not significant. We average these two values as our metallicity estimate for the system.

Interestingly, the orbit of this system is circular while at least one of the components is not rotating synchronously with the orbit. However, tidal theory suggests that the timescale for synchronous rotation is much shorter than the time scale for circularization \citep{async}, so it is unclear how this system can currently be in the state that we observe it in unless it formed in a nearly circular orbit and the system age is smaller than the synchronous rotation timescale.

 With these measurements in hand, we can now use these stars as tests of stellar models. In Figure 6 we plot the masses and radii for other low-mass eclipsing binary systems, along with the stellar models of \citep{New_Baraffe_Models_2015} for a 5 Gyr system of solar metallicity. A table of the values for stars used in this plot is available in Table \ref{MR_Conglomeration_table}. We find that each of the individual components of LP 661-13 are higher than, but marginally compatible with, the most recent stellar models. However, since in eclipsing binary observations the sum of the component radii are better determined than each individual component's radius we can investigate whether this radius sum is significantly higher than expected from stellar models. This also allows us to utilize the precision of our mass measurements in order to fully leverage the available data and test these models. We find that given the measured stellar masses of $0.308 M_\odot$ and $0.194 M_\odot$, that we expect a total radius sum of $0.510 \pm 0.005 R_\odot$. The measured radius sum, $0.536 \pm 0.006 R_\odot$ lies 5\% above the model value, implying a significant radius inflation at $4 \sigma$ confidence.

One of the most straightforward ways in which an M dwarf might be inflated is due to youth. M dwarfs take significantly longer to settle onto the main-sequence than solar type stars, and may maintain significantly larger radii for up to a billion years. If LP 661-13 is relatively young, we might expect both components to be slightly inflated. We find no evidence for Lithium in either component in our measured spectra, which can set a lower limit on their age. However, since both stars are fully convective (M $< 0.35$ M$_\odot$, \citet{structure_convection}), any Lithium originally present in the star would be convected into the interior and destroyed in approximately 10 Myr \citep{New_Baraffe_Models_2015}. Using the stellar models from \citet{New_Baraffe_Models_2015}, we find that in order to reproduce the radius sum that we measure, LP 661-13 would need to be approximately 200 Myr old. At an age of 200 Myr, we would expect an X-ray luminosity of $10^{28.5}$ erg s$^{-1}$ \citep{rebolo2000very}, which we can rule out with the ROSAT data. Furthermore, we find no detectable eccentricity in the system and the circularization timescale is 4.3 Gyr. Therefore, we find it unlikely that the average inflation of these stars can be explained through youth and that it is likely that LP 661-13 is a field age system. 

Another possible mechanism that may be responsible for M dwarfs' inflated radii is the presence of metals in the atmosphere. The presence of metals and the cool temperatures present in the atmospheres are conducive to the formation of molecules. In turn, these molecules act as a significant source of opacity in the optical, suppressing the amount of energy that is emitted in these frequency bands. To compensate for this the star emits significantly more light in the infrared than in the optical compared to a blackbody spectrum. However, it is possible that the star may also expand slightly to have a greater surface area from which to emit the energy it is producing in the stellar interior. In this case, the presence of more metals would lead to a slightly larger radius than for a star depleted in metals. While the new stellar models published by \citet{New_Baraffe_Models_2015} are not computed over a range of different metallicities, previous versions of these models \citep{Baraffe98}, found that metallicity can only account for a 3\% increase in the radius from [Fe/H] = -0.5 to [Fe/H] = 0.0. Since the average radius inflation for LP 661-13 is higher than this, it is unlikely that metallicity can fully account for the discrepancy from the stellar models, and likely there is still theoretical considerations to be addressed to fully understand the mass-radius relation of low-mass stars.

\section{Conclusions}
We present here the discovery and analysis of the eclipsing M dwarf - M dwarf binary LP 661-13. We have collected 2 years of eclipse data and precise radial velocity measurements of both components in order to obtain accurate, model-independent measurements of their masses and radii. We find that LP 661-13A is a $0.30795 \pm 0.00084$ $M_\odot$ star with a $0.3226 \pm 0.0033$ $R_\odot$ radius while LP661-13B is a $0.19400 \pm 0.00034$ $M_\odot$ star with a $0.2174 \pm 0.0023$ $R_\odot$ radius. Both components are slightly inflated in radius when compared to stellar models. However, the radius sum (which is much better constrained) is significantly (4 $\sigma$) inflated when compared to the expected radius sum from stellar models. Because the orbit of the system is circularized, it is unlikely that this inflation can be explained through youth. Metallicity is also insufficient to explain the total radius inflation we observe. 

In the future, the most straightforward way to improve the measurements of this system is to continue out of eclipse monitoring and obtain additional eclipse observations. Because we have observed some spot evolution between observing seasons, additional evolution will help to break the degeneracies between starspot coverage and inferred stellar radii in the model and provide better constraints on the fundamental parameters of this system. Additionally, we can potentially probe the origin of the radius inflation by investigating the marginal X-ray activity as seen by ROSAT and attempt to measure its surface magnetic field. Eclipse measurements in other photometric bandpasses will allow us, with the trigonometric parallax distance we have in hand, to measure the effective temperatures of each component as well, which will serve as another test of stellar models. LP 661-13 represents another low-mass stellar test case measured with high accuracy and will be a useful benchmark for current and future stellar models. LP 661-13 is positioned equatorially on the sky and therefore is a good object for further study from both northern and southern facilities.

\section*{Acknowledgments}

The MEarth Team gratefully acknowledges funding from the David and Lucille Packard Fellowship for Science and Engineering (awarded to D.C.). This material is based upon work supported by the National Science Foundation under grants AST-0807690, AST-1109468, and AST-1004488 (Alan T. Waterman Award). This publication was made possible through the support of a grant from the John Templeton Foundation. The opinions expressed in this publication are those of the authors and do not necessarily reflect the views of the John Templeton Foundation. E.R.N. was supported by the NSF Graduate Research Fellowship. This research has made extensive use of NASAs Astrophysics Data System (ADS), and the SIMBAD database, operated at CDS, Strasbourg, France.

\clearpage

\bibliography{ms}

\clearpage

\begin{table}
\begin{center}
\caption{TRES Radial Velocities for LP 661-13 (Barycentric)}
\label{RV_table}
\begin{tabular}{crrrr} 
\tableline\tableline
Date (HJD) & $v_1$ (km s$^{-1}$) & $v_2$ (km s$^{-1}$) \\
\tableline
\tableline 
2456706.708137 & -38.974 & 61.767 \\
2456709.709816 & 26.444 & -41.731 \\
2456738.656903 & -8.507 & 13.448 \\
2456740.665846 & -9.123 & 14.699 \\
2456741.701681 & 35.657 & -56.743\\
2456742.704103 & 23.730 & -37.638\\
2456744.696853 & -35.311 & 56.147 \\
2457018.893616 & 23.966 & -38.260 \\
2457019.845650 & 36.549 & -58.105 \\
2457020.921800 & -8.778 & 14.120 \\
2457025.844636 & -19.396 & 30.542 \\
2457026.835799 & -37.595  & 59.583 \\
2457028.858941 & 38.496 & -61.060 \\
\tableline
\end{tabular}
\end{center}
\end{table}

\newpage

\begin{table}
\begin{center}
\caption{MEarth Photometry for LP 661-13}
\label{Photometry_table}
\begin{tabular}{crrrrr} 
\tableline\tableline
Date (HJD) & $\Delta$ magnitude & error (mag) & telescope \\
\tableline
\tableline 
2456685.55228 & -0.0012 & 0.0020 & tel15 \\
2456685.55287 & -0.0003 & 0.0020 & tel15 \\
2456685.56806 & -0.0006 & 0.0019 & tel15 \\
2456685.56866 & -0.0014 & 0.0019 & tel15 \\
2456685.58610 & -0.0031 & 0.0019 & tel15 \\
... & ... & ... & ... \\
\tableline
\end{tabular}
\end{center}
\end{table}

%\begin{singlespace}
\begin{sidewaystable}
\begin{center}
\caption{Model Parameters}
\label{model_table}
\begin{tabular}{crrrrr} 
\tableline\tableline
\centering
Parameter & Value & Prior\\
\tableline
\tableline 
$J_{\textrm{MEarth}}$ & Varied & Uniform \\
$(R_1 + R_2) / a$ & Varied & Uniform \\
$R_2 / R_1$ & Varied & Uniform \\
cos $i$ & Varied & Uniform (isotropic in $i$) \\
$M_2 / M_1$ & Varied & Uniform \\
$K_1 + K_2$ & Varied & Uniform \\
$u_{1,1}$ & Varied & Uniform ($0.2232 - 0.2728$) \\
$u_{1,2}$ & Varied & Uniform ($0.5364 - 0.6556$) \\
$u_{2,1}$ & Varied & Uniform ($0.4581 - 0.5599$) \\
$u_{2,2}$ & Varied & Uniform ($0.3717 - 0.4543$) \\
$\beta_1$ & 0.32 & ... \\
$\beta_2$ & 0.32 & ... \\
$L_3$ & 0 & ... \\
$F$ & Varied & Uniform \\
$f_1$ & Varied (Fixed for final solution) & Uniform \\
$f_2$ & Varied (Fixed for final solution) & Uniform \\
$a$ & Varied & Uniform \\
$b$ & Varied & Uniform \\
$e \textrm{cos} \omega$ & Varied (Fixed for final solution) & Uniform \\
$e \textrm{sin} \omega$ & Varied (Fixed for final solution) & Uniform \\
$P$ & Varied & Uniform \\
$T_0$ & Varied & Uniform \\
$\gamma$ & Varied & Uniform \\
\tableline
\end{tabular}
\end{center}
\end{sidewaystable}
%\end{singlespace}

%\begin{singlespace}
\begin{sidewaystable}
\begin{center}
\begin{tabular}{crr} 
Table 3 (Continued):\\
\tableline\tableline
\centering
Description & \\
\tableline
\tableline 
Central surface brightness ratio (secondary/primary) in $MEarth$ \\
Sum of the radius of each component divided by the semi major axis \\
Radius ratio \\
Cosine of the orbital inclination  \\
Mass ratio  \\
Sum of RV semi-amplitudes (km/s) \\
Linear limb-darkening coefficient for primary \\
Quadratic limb-darkening coefficient for primary \\
Linear limb-darkening coefficient for secondary \\
Quadratic limb-darkening coefficient for secondary \\
Gravity darkening exponent for primary \\
Gravity darkening exponent for secondary \\
Third light component divided by total system light \\
Ratio of rotational to orbital frequency \\
Fraction of spots eclipsed \\
Base spottedness of star \\
Out-of-eclipse sine coefficient \\
Out-of-eclipse cosine coefficient \\
Eccentricity times cosine of argument of periastron \\
Eccentricity times sine of argument of periastron \\
Orbital period of system (heliocentric) (days) \\
Epoch of primary eclipse (HJD - 2456600.0) \\
Barycentric systemic radial velocity of system (km/s) \\
\tableline
\end{tabular}
\end{center}
\end{sidewaystable}
%\end{singlespace}

\newpage

\begin{table}
\begin{center}
\caption{Parameters for Model With Starspots on Primary Component}
\label{results_table_prim}
\begin{tabular}{crrrr} 
\tableline\tableline
\centering
Parameter & 2014 Season & 2015 Season \\
\tableline
\tableline 
$J_{\textrm{MEarth}}$ & $0.8705^{+0.0298}_{-0.0268}$ & $0.8698^{+0.0261}_{-0.0177}$ \\
$(R_1 + R_2) / a$ & $0.0574^{+0.0007}_{-0.0008}$ & $0.0574^{+0.0005}_{-0.0004}$\\
$R_2 / R_1$ & $0.6836^{+0.1065}_{-0.1539}$ & $0.7500^{+0.0674}_{-0.0717}$\\ 
cos $i$ & $0.0425^{+0.0008}_{-0.0013}$ & $0.0426^{+0.0005}_{-0.0006}$ \\
$M_2 / M_1$ & $0.6299 \pm 0.0020$ & $0.6299 \pm 0.0020$ \\
$K_1 + K_2$ & $100.983 \pm 0.077$ & $100.983 \pm 0.077$ \\
$u_{1,1}$ & $0.2474^{+0.0188}_{-0.0189}$ & $0.2579^{+0.0105}_{-0.0269}$ \\
$u_{1,2}$ & $0.6102^{+0.0376}_{-0.0608}$ & $0.6413^{+0.0105}_{-0.0218}$ \\
$u_{2,1}$ & $0.4901^{+0.0530}_{-0.0272}$ & $0.4723^{+0.0305}_{-0.0116}$  \\
$u_{2,2}$ & $0.4123^{+0.0324}_{-0.0299}$ & $0.4157^{+0.0292}_{-0.0306}$\\
$F$ & $1.2288^{+0.0030}_{-0.0145}$ & $1.2284^{+0.0039}_{-0.0160}$\\
$f_1$ & $0.0684^{+0.2328}_{-0.0473}$ & $0.0561^{+0.1512}_{-0.0353}$\\
$f_2$ & $0.0157^{+0.0838}_{-0.0110}$ & $0.0104^{+0.0212}_{-0.0070}$\\
$a$ & $0.0009^{+0.0005}_{-0.0007}$ & $0.00035^{+0.00026}_{-0.00035}$\\
$b$ & $-0.0003^{+0.0003}_{-0.0002}$ & $-0.00016^{+0.00027}_{-0.00023}$\\
$e \textrm{cos} \omega$ & $-0.00003^{+0.00003}_{-0.00006}$ & $0.000000^{+0.000017}_{-0.000038}$\\
$e \textrm{sin} \omega$ & $-0.000003^{+0.000450}_{-0.000090}$ & $-0.000003^{+0.000038}_{-0.000054}$\\
$P$ & $4.704364^{+0.000040}_{-0.000008}$ & $4.704360^{+0.000005}_{-0.000005}$\\
$T_0$ & $105.5602^{+0.0001}_{-0.0006}$ & $105.5598^{+0.0003}_{-0.0003}$\\
$\gamma$ & $-0.009 \pm 0.014$ & $-0.009 \pm 0.014$ \\
\tableline
\end{tabular}
\end{center}
\end{table}

\begin{table}
\begin{center}
\caption{Model Parameters for Starspots on Secondary Component}
\label{results_table_sec}
\begin{tabular}{crrrr} 
\tableline\tableline
\centering
Parameter & 2014 Season & 2015 Season  \\
\tableline
\tableline 
$J_{\textrm{MEarth}}$ & $0.9114^{+0.0453}_{-0.0294}$ & $0.8729^{+0.0254}_{-0.0252}$ \\
$(R_1 + R_2) / a$ & $0.0580^{+0.0012}_{-0.0010}$ & $0.0573^{+0.0006}_{-0.0006}$\\
$R_2 / R_1$ & $0.6894^{+0.1558}_{-0.0929}$ & $0.7493^{+0.1297}_{-0.0722}$\\ 
cos $i$ & $0.0430^{+0.0015}_{-0.0014}$ & $0.0424^{+0.0008}_{-0.0007}$\\
$M_2 / M_1$ & $0.6299 \pm 0.0020$ & $0.6299 \pm 0.0020$ \\
$K_1 + K_2$ & $100.983 \pm 0.077$ & $100.983 \pm 0.077$ \\
$u_{1,1}$ & $0.2393^{+0.0230}_{-0.0116}$ & $0.2582^{+0.0100}_{-0.0167}$\\
$u_{1,2}$ & $0.5777^{+0.0469}_{-0.0316}$ & $0.6395^{+0.0122}_{-0.0361}$\\
$u_{2,1}$ & $0.5048^{+0.0379}_{-0.0352}$ & $0.4767^{+0.0406}_{-0.0156}$\\
$u_{2,2}$ & $0.4034^{+0.0331}_{-0.0239}$ & $0.3927^{+0.0357}_{-0.0161}$\\
$F$ & $1.778^{+0.0102}_{-0.4461}$ & $0.9098^{+0.2803}_{-0.0052}$\\
$f_1$ & $0.0193^{+0.2414}_{-0.0182}$ & $0.0023^{+0.0075}_{-0.0016}$\\
$f_2$ & $0.0154^{+0.0850}_{-0.0144}$ & $0.0024^{+0.0105}_{-0.0015}$\\
$a$ & $0.0033^{+0.0019}_{-0.0032}$ & $0.0017^{+0.0037}_{-0.0021}$\\
$b$ & $0.0010^{+0.0006}_{-0.0009}$ & $0.00010^{+0.0038}_{-0.0039}$\\
$e \textrm{cos} \omega$ & $-0.000058^{+0.000041}_{-0.000067}$ & $-0.000055^{+0.000031}_{-0.000029}$\\
$e \textrm{sin} \omega$ & $-0.00000^{+0.00181}_{-0.00069}$ & $0.000013^{+0.00012}_{-0.000029}$\\
$P$ & $4.70438^{+0.00002}_{-0.00001}$ & $4.704362^{+0.0000051}_{-0.0000052}$\\
$T_0$ & $105.56034^{+0.00006}_{-0.00006}$ & $105.55972^{+0.00033}_{-0.00034}$\\
$\gamma$ & $-0.009 \pm 0.014$ & $-0.009 \pm 0.014$ \\
\tableline
\end{tabular}
\end{center}
\end{table}

%\begin{singlespace}
\begin{sidewaystable}
\begin{center}
\caption{Joint Model Parameters$^{a}$}
\label{Final_Model_Params}
\begin{tabular}{crrrr} 
\tableline\tableline
\centering
Parameter & Value (Prim. spots, L$_3$ = 0) & Value (Prim. spots, L$_3$ = 0.01)  \\
\tableline
\tableline 
$J_{\textrm{MEarth}}$ & $0.9004^{+0.0037}_{-0.0085}$ & $0.8933^{+0.0086}_{-0.0209}$ \\
$(R_1 + R_2) / a$ & $0.05714^{+0.00037}_{-0.00035}$ & $0.05710^{+0.00045}_{-0.00038}$ \\
$R_2 / R_1$ & $0.6745^{+0.0095}_{-0.0112}$ & $0.675^{+0.038}_{-0.013}$ \\
cos $i$ & $0.04206^{+0.00034}_{-0.00038}$ & $0.04202^{+0.00040}_{-0.00045}$ \\
$M_2 / M_1$ & $0.634^{+0.020}_{-0.006}$ & $0.6314^{+0.0082}_{-0.0080}$ \\
$K_1 + K_2$ & $100.96^{+0.14}_{-0.18}$ & $100.834918085^{+0.10}_{-0.11}$ \\
$u_{1,1}$ & $0.2411^{+0.0040}_{-0.0046}$ & $0.2419^{+0.0117}_{-0.0038}$ \\
$u_{1,2}$ & $0.5770^{+0.0085}_{-0.0082}$ & $0.5796^{+0.0098}_{-0.0167}$ \\
$u_{2,1}$ & $0.5113^{+0.0055}_{-0.0081}$ & $0.498^{+0.013}_{-0.025}$ \\
$u_{2,2}$ & $0.4140^{+0.0044}_{-0.0145}$ & $0.4151^{+0.0081}_{-0.0120}$ \\
$F$ (season 1) & $1.226^{+0.010}_{-0.012}$ & $1.205^{+0.026}_{-0.023}$ \\
$f_1$ (season 1) & 1 (fixed) & 1 (fixed) \\
$f_2$ (season 1) & 0 (fixed) & 0 (fixed) \\
$a$ (season 1) & $0.00137^{+0.00013}_{-0.00002}$ & $0.001351^{+0.000025}_{-0.000073}$ \\
$b$ (season 1) &$-0.000492^{+0.000006}_{-0.000019}$ & $-0.000499^{+0.000009}_{-0.000082}$ \\
$F$ (season 2)& $1.2300^{+0.0008}_{-0.0159}$ & $1.2301^{+0.0007}_{-0.0015}$ \\
$f_1$ (season 2) & 1 (fixed) & 1 (fixed) \\
$f_2$ (season 2) & 0 (fixed) & 0 (fixed) \\
$a$ (season 2) & $0.00134^{+0.00002}_{-0.00014}$ & $0.00134^{+0.00002}_{-0.00011}$ \\
$b$ (season 2) & $-0.000491^{+0.000057}_{-0.000006}$ & $-0.000484^{+0.000043}_{-0.000017}$ \\
$e \textrm{cos} \omega$ & 0 (fixed) & 0 (fixed)  \\
$e \textrm{sin} \omega$ & 0 (fixed) & 0 (fixed) \\
$P$ & $4.7043512^{+0.0000013}_{-0.0000010}$ & $4.7043504^{+0.0000017}_{-0.0000014}$ \\
$T_0$ & $105.560353^{+0.000038}_{-0.000040}$ & $105.560446^{+0.000094}_{-0.000087}$ \\
$\gamma$ & $-0.009 \pm 0.014$ & $-0.009 \pm 0.014$ \\
\tableline
\end{tabular}
\end{center}
\end{sidewaystable}
%\end{singlespace}
\clearpage

%\begin{singlespace}
\begin{sidewaystable}
\begin{center}
\begin{tabular}{crrr} 
Table 6 (Continued)\\
\tableline\tableline
\centering
Value (Sec. spots, L$_3$ = 0) & Value (Sec. spots, L$_3$ = 0.01) \\
\tableline
\tableline 
$0.891^{+0.015}_{-0.017}$ & $0.891^{+0.012}_{-0.024}$ \\
$0.05730^{+0.00035}_{-0.00031}$ & $0.05710^{+0.00053}_{-0.00031}$\\
$0.679^{+0.033}_{-0.018}$ & $0.678^{+0.028}_{-0.010}$\\ 
$0.04218^{+0.00039}_{-0.00037}$ & $0.04189^{+0.00054}_{-0.00023}$\\
$0.6320^{+0.0081}_{-0.0048}$ & $0.632^{+0.012}_{-0.013}$\\
$100.94^{+0.12}_{-0.13}$ & $101.54^{+0.15}_{-0.12}$\\
$0.242^{+0.019}_{-0.010}$ & $0.2429^{+0.0070}_{-0.0055}$\\
$0.582^{+0.046}_{-0.028}$ & $0.596^{+0.037}_{-0.029}$\\
$0.511^{+0.020}_{-0.023}$ & $0.512^{+0.007}_{-0.026}$\\
$0.411^{+0.014}_{-0.022}$ & $0.405^{+0.011}_{-0.013}$\\
$1.000^{+0.087}_{-0.094}$ & $0.995^{+0.032}_{-0.044}$\\
1 (fixed) & 1 (fixed) \\
0 (fixed) & 0 (fixed) \\
$0.0001002^{+0.0000049}_{-0.0000035}$ & $0.0001000^{+0.0000040}_{-0.0000025}$\\
$0.0001005^{+0.0000043}_{-0.0000029}$ & $0.0001024^{+0.0000045}_{-0.0000030}$\\
$1.249^{+0.057}_{-0.029}$ & $1.230^{+0.035}_{-0.024}$\\
1 (fixed) & 1 (fixed) \\
0 (fixed) & 0 (fixed) \\
$0.00136^{+0.00014}_{-0.00005}$ & $0.001351^{+0.000036}_{-0.000047}$\\
$-0.000495^{+0.000016}_{-0.000030}$ & $-0.000485^{+0.000033}_{-0.000016}$\\
0 (fixed) & 0 (fixed) \\
0 (fixed) & 0 (fixed) \\
$4.7043515^{+0.0000012}_{-0.0000017}$ & $4.7043516^{+0.0000014}_{-0.0000016}$\\
$105.560361^{+0.000075}_{-0.000069}$ & $105.560371^{+0.000086}_{-0.000072}$\\
$0.03 \pm 0.15 $ & $-0.03^{+0.14}_{-0.10}$\\
\tableline
\end{tabular}
\\a) We believe that star spot must be located on the primary component due to the behavior of F between seasons. We utilize L$_3$ = 0.01 as an upper limit on the third light contamination and it is likely that L$_3$ is much less than 0.01. Of these four models, we believe the first column (Prim. spots, L$_3$ = 0) to be the most likely scenario.
\end{center}
\end{sidewaystable}
%\end{singlespace}
\clearpage

\begin{table}
\begin{center}
\caption{Physical parameters of LP 661-13}
\label{Final_System_Parameters_table}
\begin{tabular}{crrr} 
\tableline\tableline
Parameter & Value & Source\\
\tableline
\tableline 
$M_1$ ($M_\odot$) & $0.30795 \pm 0.00084$ & This work \\
$M_2$ ($M_\odot$) & $0.19400 \pm 0.00034$ & This work  \\
$R_1$ ($R_\odot$) & $0.3226 \pm 0.0033$ & This work  \\
$R_2$ ($R_\odot$) & $0.2174 \pm 0.0023$ & This work  \\
$V$ & 14.03 & \citet{reid} \\
$R$ & 12.75 & \cite{reid} \\
$I$ & 11.17 & \cite{reid} \\
$J$ & $9.63 \pm 0.02$  & \citet{2mass} \\
$H$ & $9.07 \pm 0.02$ & \citet{2mass} \\
$K_s$ & $8.76 \pm 0.02$ & \citet{2mass} \\
Spectral Type & M3.5 & \citet{reid} \\
NIR Spectral Type & M4.27 & \cite{Terrien} \\
Distance (pc) & $24.9 \pm 1.3$  & This work \\
Proper motion (RA, mas yr$^{-1}$) & $75.1 \pm 2.0$ & This work  \\
Proper motion (DEC, mas yr$^{-1}$) & $-227.3 \pm 2.0$ & This work  \\
$\gamma$ (km/s) & $-0.009 \pm 0.014$ & This work \\
Period (days) & $4.7043512^{+0.0000013}_{-0.0000010}$ & This work \\
Epoch of primary eclipse (HJD) & $2456705.560353^{+0.000038}_{-0.000040}$ & This work \\
$[$Fe/H$]$ & -0.07 & This work \\
\tableline
\end{tabular}
\end{center}
\end{table}

\begin{table}
\begin{center}
\caption{Measured Masses and Radii for Low-Mass Stars}
\label{MR_Conglomeration_table}
\begin{tabular}{crrrr} 
\tableline\tableline
Star & Mass (M$_\odot$) & Radius (R$_\odot$) & Source\\
\tableline
CMDraA & $0.2310 \pm 0.0009$ & $0.2534 \pm 0.0019$ & \citet{Morales09} \\
CMDraB & $0.2141 \pm 0.0010$ & $0.2396 \pm 0.0015$ & \citet{Morales09} \\
CU CncB & $0.3980 \pm 0.0014$ & $0.3908 \pm 0.0094$ & \citet{ribas2003} \\
GJ3236A & $0.376 \pm 0.016$ & $0.3795 \pm 0.0084$ & \citet{2009ApJ...701.1436I} \\
GJ3236B & $0.281 \pm 0.015$ & $0.300 \pm 0.015$ & \citet{2009ApJ...701.1436I} \\
HATS551-027A & $0.244 \pm 0.003$ & $0.261 \pm 0.006$ & \citet{Zhou} \\
HATS551-027B & $0.179 \pm 0.002$ & $0.216 \pm 0.010$ & \citet{Zhou} \\
KOI126B & $0.2413 \pm 0.0030$ & $0.2543 \pm 0.0014$ & \citet{carter2011} \\
KOI126C & $0.2127 \pm 0.0026$ & $0.2318 \pm 0.0013$ & \citet{carter2011} \\
LSPMJ1112+7626A & $0.395 \pm 0.002$ & $0.3814 \pm 0.0028$ & \citet{Irwin_41day} \\
LSPMJ1112+7626B & $0.275 \pm 0.001$ & $0.3000 \pm 0.0045$ & \citet{Irwin_41day} \\
LP133-373A & $0.340 \pm 0.014$ & $0.33 \pm 0.02$ & \citet{vaccaro2007} \\
LP133-373B & $0.340 \pm 0.014$ & $0.33 \pm 0.02$ & \citet{vaccaro2007} \\
LP661-13A & $0.30795 \pm 0.00084$ & $0.3226 \pm 0.0033$ & This work \\
LP661-13B & $0.19400 \pm 0.00034$ & $0.2174 \pm 0.0023$ & This work \\
MG1-2056316B & $0.382 \pm 0.001$ & $0.374 \pm 0.002$ & \citet{Kraus2011} \\
1RXSJ154727.5+450803A & $0.2576 \pm 0.0085$ & $0.2895 \pm 0.0068$ & \citet{hartman} \\
1RXSJ154727.5+450803B & $0.2585 \pm 0.0080$ & $0.2895 \pm 0.0068$ & \citet{hartman} \\
SDSS-MEB-1A & $0.272 \pm 0.020$ & $0.268 \pm 0.009$ & \citet{blake2008} \\
SDSS-MEB-1B & $0.240 \pm 0.022$ & $0.248 \pm 0.0084$ & \citet{blake2008} \\
WTS19g-4-02069A & $0.53 \pm 0.02$ & $0.51 \pm 0.01$ & \citet{nefs2013} \\
WTS19g-4-02069B & $0.143 \pm 0.006$ & $0.174 \pm 0.006$ & \citet{nefs2013}\\
\tableline
\tableline 
\tableline
\end{tabular}
\end{center}
\end{table}

\newpage

\clearpage
\clearpage

\clearpage

\begin{figure}
\label{massradius}
\centering
\includegraphics[width=0.85\linewidth]{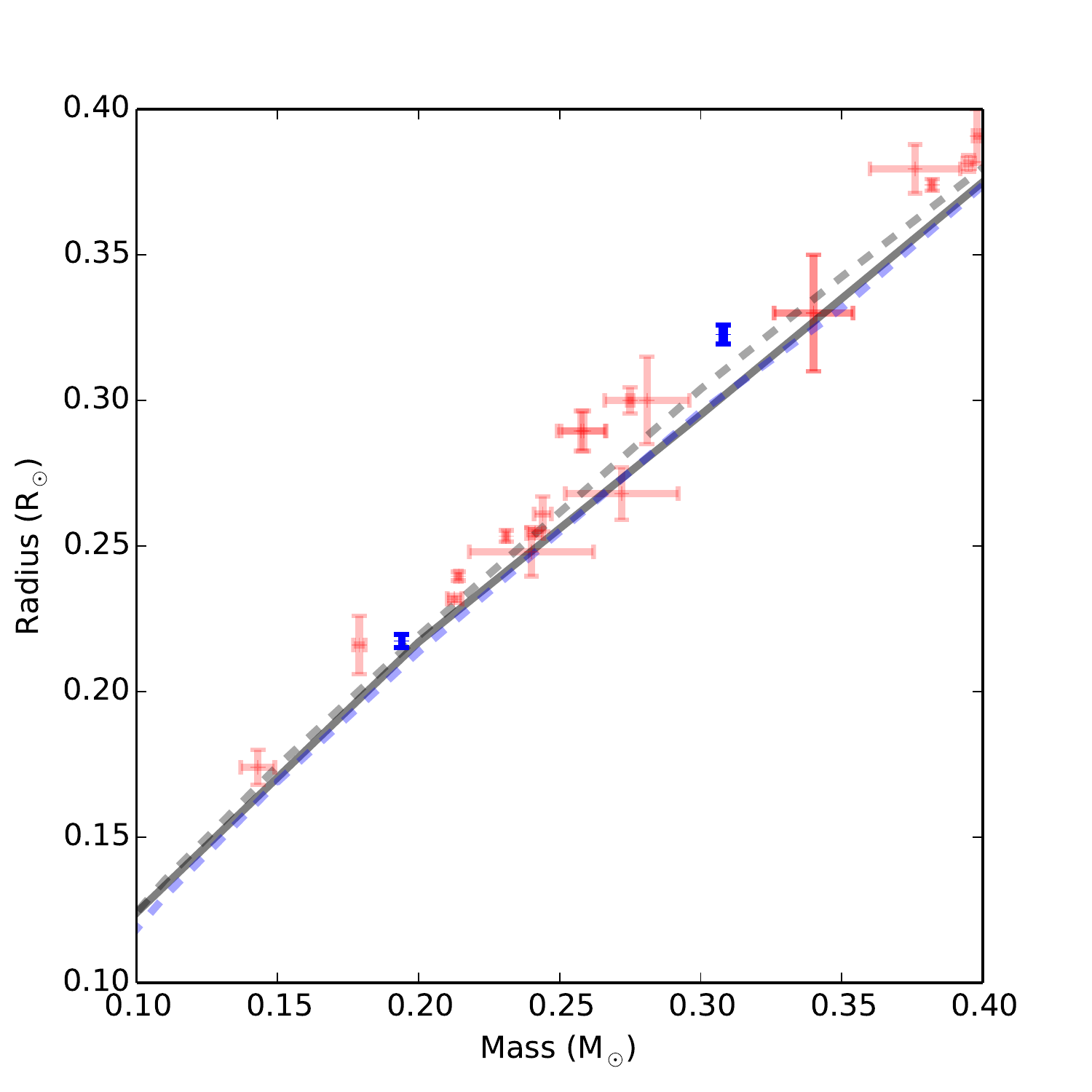} \\
%\begin{singlespace}
\caption{The masses and radii for nearby double lined low-mass eclipsing binary stars with precise measurements (see Table \ref{MR_Conglomeration_table}). The fully convective boundary is at $0.35$ M$_\odot$. We plot SDSS-MEB-1 \citep{blake2008}, GJ 3236 \citep{2009ApJ...701.1436I}, CMDra \citep{Morales09}, LP 133-373 \citep{vaccaro2007}, MG1-2056316 \citep{Kraus2011}, KOI-126 \citep{carter2011}, CU Cnc \citep{ribas2003}, 1RXSJ154727 \citep{hartman}, HATS551-027 \citep{Zhou}, LSPM J1112+7626 \citep{Irwin_41day}, and WTS19g-4-02069 \citep{nefs2013} in red. LP661-13 A and B are indicated by the dark blue crosses. The black line is the stellar model from \citet{New_Baraffe_Models_2015} for a 5 Gyr system with solar metallicity. The black dashed line is for a solar metallicity system with an age of 10 Gyr from the models of \citet{structure_convection} while the dashed blue line is for the 10 Gyr, [Fe/H] = -0.5 model from \citet{structure_convection}. We find that in aggregate the stellar models tend to underpredict the radius for a star of a given mass. While the radius  for each component of LP 661-13 is marginally consistent with the stellar models, we find that the much better constrained radius sum is significantly inflated compared to that predicted by the stellar model.}
%\end{singlespace}
\end{figure}

\end{document}